\documentclass[preprint]{aastex7}
\newcommand{\elleff}{\Tilde{\ell}}
\newcommand{\velh}{\vel_{\rm h}}
\newcommand{\teff}{T_{\rm eff}}
\newcommand{\Rcz}{R_{\rm CZ}}
\newcommand{\vel}{{\rm v}}
\newcommand{\velvec}{{\rm \mathbf v}}
\newcommand{\rin}{r_{\rm in}}
\newcommand{\rout}{r_{\rm out}}

\newcommand{\Virms}{\vel_\mathrm{i,rms}}
\newcommand{\dr}[1]{\frac{\mathrm{d}#1}{\mathrm{d}r}}
\newcommand{\AvgS}[1]{\left< {#1} \right>_{\mathcal{S}}}
\newcommand{\AvgST}[1]{\left< {#1} \right>_{\mathcal{S},t}}
\newcommand{\AvgT}[1]{\left< {#1} \right>_t}

\newcommand{\Rstar}{R_{\rm star}}
\newcommand{\velsurf}{\vel_{\rm surf}}
\newcommand{\velsurfext}{\vel_{\rm surf, ro98}}

\usepackage{amsmath}
\usepackage{newtxmath}

\begin{document}

\title{A core-sensitive mixed $f$/$g$ mode of the Sun predicted by wave topology and hydrodynamical simulation}

\author[orcid=0000-0002-6686-7956,gname=Arthur,sname='Le Saux']{Arthur Le Saux}
\altaffiliation{These authors contributed equally to this work}
\affiliation{Université Paris-Saclay, Université Paris Cité, CEA, CNRS, AIM, Gif-sur-Yvette, F-91191, France.}
\affiliation{Laboratoire de Météorologie Dynamique / Institut Pierre-Simon Laplace (LMD/IPSL), Sorbonne Université, Centre National de la Recherche Scientifique (CNRS), École Polytechnique, École Normale Supérieure (ENS)}
\email[show]{arthur.lesaux@cea.fr}  

\author[orcid=0000-0001-6136-4164,gname=Armand, sname='Leclerc']{Armand Leclerc} 
\altaffiliation{These authors contributed equally to this work}
\affiliation{ENS de Lyon, CRAL UMR5574, Universite Claude Bernard Lyon 1, CNRS, Lyon, F-69007, France}
\email{armand.leclerc@ens-lyon.fr}

\author[gname=Guillaume,sname=Laibe]{Guillaume Laibe}
\affiliation{ENS de Lyon, CRAL UMR5574, Universite Claude Bernard Lyon 1, CNRS, Lyon, F-69007, France}
\affiliation{Institut Universitaire de France}
\email{glaibe@ens-lyon.fr}

\author[gname=Pierre,sname=Delplace]{Pierre Delplace}
\affiliation{ENS de Lyon, CNRS, Laboratoire de physique, F-69342 Lyon, France}
\email{pierre.delplace@ens-lyon.fr}

\author[sname=Antoine,gname=Venaille]{Antoine Venaille}
\affiliation{ENS de Lyon, CNRS, Laboratoire de physique, F-69342 Lyon, France}
\email{antoine.venaille@ens-lyon.fr}

\begin{abstract}
Helioseismology has revolutionized our understanding of the Sun by analyzing its global oscillation modes. 
However, the solar core remains elusive, limiting a full understanding of its evolution.
In this work, we study a previously unnoticed global oscillation mode of the Sun using a fully compressible, hydrodynamical simulation of the solar interior, and assess that it is a mixed $f$/$g$ mode with a period of about one hour. This is the first global stellar hydrodynamics simulation that successfuly couple compressible and gravity modes.
To understand this coupling, we invoke a recent theory on the nature of $f$-modes seen through the prism of wave topology, characterizing their ability to propagate deep into stellar interiors. We demonstrate that the mixed $f$/$g$ mode is highly sensitive to the core's rotation rate, providing a new promising pathway to explore the Sun's core.
\end{abstract}

\keywords{}


\section{Introduction} 

The Sun, our closest star, serves as a fundamental reference for understanding stellar evolution. Its proximity offers a unique window into the physical processes governing the life cycles of stars, with broader implications for astrophysics, as stars are the building blocks of galaxies and hosts of exoplanets. 
In recent decades, helioseismology - the science that studies solar oscillations - has revolutionized our understanding of the Sun.
Seismic inversion techniques of these modes have provided precise constraints of the Sun’s internal structure and dynamics, such as the location of the interface between the radiative and convective zones, the abundance of helium in the convective envelope, the efficiency of chemical diffusion or the rotation profile of the solar interior \citep{Howe2009,Basu2016}. These advances have been driven by high-precision data from space telescopes \citep{gabriel1997,hoeksema2018} and ground-based networks of solar observatories \citep{fossat1991,chaplin1996,harvey1996}. Today, we know most of the Sun's internal structure (outer 90 \% of the total radius) and rotation profile (outer 80 \%) \citep[see the detailed review by][]{JCD2021}. However, a global understanding of the theories of angular momentum transport and mixing of elements in stellar interiors requires observational constraints for the innermost 10-20\% of the Sun's radial layers \citep{Leibacher2023}.

Unfortunately, the solar core remains inaccessible to helioseismology. Indeed, the non-radial acoustic modes, or $p$-modes, used for inversion are refracted before reaching the core, where their horizontal wavelength becomes infinitesimally small, confining them outside. 
Recently, inertial and Rossby waves, supported by solar rotation, have provided new insights into superadiabaticity and turbulent viscosity in the deep convection zone \citep{Gizon2021}, and on the Sun's latitudinal differential rotation \citep{Bekki2024}. But these modes are confined to the convective envelope and do not probe the core. 

So far, the only constraints on the solar core come from neutrinos, which have recently revealed the Carbon-Nitrogen-Oxygen cycle and provided insights into stellar metallicity and energy production \citep{Borexino2020}.  A long-standing goal of helioseismology is the detection of solar internal gravity waves, or $g$-modes, buoyant oscillations that could provide unprecedented information about the Sun's interior. 
While some claims of detection exist \citep{Garcia2007, Fossat2017}, these results remain unconfirmed \citep{appourchaux2010, Schunker2018}. Their detection is difficult due to their small amplitude at the surface, and frequency domain overlapping with solar granulation noise \citep{belkacem2022}, making them elusive to current instruments.

In this study, we show that some $g$-modes couple with $f$-modes at large horizontal wavelength and form mixed $f$/$g$ modes, which have amplitude both in the core and at the surface. This key property makes them potential observational probes of the Sun’s deep interior dynamics. This result, evidenced by state-of-the-art hydrodynamical simulations, challenge the usual depiction of $f$-modes as being surface waves \citep{deubner1984,gough1993,aerts2010}, a description which only holds at short horizontal wavelength. We demonstrate that at large horizontal wavelength, $f$-modes are topological waves that propagate in the bulk of the star, as recently predicted by \citet{leclerc2022}. Topological waves are a class of large-scale modes characterized by topological arguments inherited from condensed matter physics \citep{delplace2017,perrot2019,perez2022topological}. The identification of such topological $f$-modes explains their ability to couple with $g$-modes, as they have spatial and frequency overlaps.

\section{The two kinds of $f$-modes}

The normal modes of non-rotating non-magnetic stars are classified as $p$-modes, $g$-modes or $f$-modes as by the seminal paper \citet{cowling1941}. Each mode is identified by two numbers: its radial order $n$ and angular degree $\ell \geq 0$ , which correspond to the numbers of radial and angular nodes of the perturbation of pressure respectively. The $p$-modes, characterized by high frequencies, permeate the entire solar interior, whereas the $g$-modes, with lower frequencies, are confined to the radiative region. The $f$-modes are the $n=0$ modes and as such are fundamental modes with generically intermediate frequencies \citep{unno1979}. In the literature, $f$-modes have mostly been described as motion of the free surface of the Sun as surface gravity waves in the plane-parallel approximation \citep{deubner1984,gough1993}, which works for short horizontal wavelengths with $\ell \gg 1$ and in good agreement with numerical models and helioseismic data \citep{christensen2002,Basu2016}. These surface waves are mostly unaffected by the Sun's interior and cannot provide information about its internal structure. Instead, they have been used to measure the solar radius and its variations \citep{antia1998,lefebvre2007}.\\
Nevertheless, the interpretation of $f$-modes as surface waves does not hold at low $\ell$. Recent studies show that for long horizontal wavelengths ($\ell \sim 1$), $f$-modes are Lamb-like waves \citep{perrot2019,leclerc2022}. These waves have been named due to similarities with the atmospheric Lamb waves \citep{lamb1911,bretherton1969}. While atmospheres support Lamb waves trapped at their bottom solid boundary, Lamb-like waves in stars occur at a specific radius, depending on sphericity. This radius is defined at the location where the characteristic frequency
\begin{equation}
S = \frac{c_\mathrm{s}}{2g}\left( N^2 - \frac{g^2}{c_\mathrm{s}^2} \right) - \frac{1}{2}\frac{\mathrm{d}c_\mathrm{s}}{\mathrm{d}r} + \frac{c_\mathrm{s}}{r} ,
\label{eq:stellarS}
\end{equation}
goes to zero and changes sign. In this equation $g$ is the gravity field, $c_\mathrm{s}$ is the sound speed, and $N$ is the buoyancy frequency. Therefore, $f$-modes at low degree $\ell$ are Lamb-like waves, and propagate close to the radius where $S=0$. In spherical geometry, the curvature term $c_\mathrm{s}/r$ in Eq.~\eqref{eq:stellarS} ensures that the frequency $S$ will change sign at least once in any star. This property was never identified before, and is key to understanding why $f$-modes leave the surface and hybridize with $g$-modes.

The prediction of stellar Lamb-like waves relies on an analysis of the topological properties of the propagation equation of local plane waves \citep{leclerc2022}. In summary, this topological analysis shows that the region where $S=0$ is a phase singularity for plane waves, which results in a branch of modes, here the Lamb-like modes, whose frequencies transition from the low-frequency ($g$-modes) to the high-frequency ($p$-modes) wavebands as $\ell$ increases. These modes, classified as topological modes, are imposed to localize where $S=0$ as predicted by the bulk-boundary correspondence. For further details on wave topology, we refer the reader to \citep{bernevig2013, perrot2019, delplace2022,perez2022topological}. \\

Then, as $\ell$ increases, the $f$-modes is confined in outer layers, and become the surface wave usually described. These two types of waves forming the $f$ branch - Lamb-like and surface - arise from competing effects between the buoyant-acoustic $S$ and the Lamb frequencies, the latter being defined as $L_\ell = \sqrt{\ell(\ell+1)}c_\mathrm{s}/r$. As shown in Supplementary Materials Eq.~\eqref{eq:schro}, the profile of $S$ tends to trap low $\ell$ modes where it is zero, while $L_\ell$ tends to repulse high-frequency modes to the surface. The competition between the two effects place the $f$-modes in different regions of the star depending on the value of $\ell$.

\begin{figure*}
    \centering
    \plotone{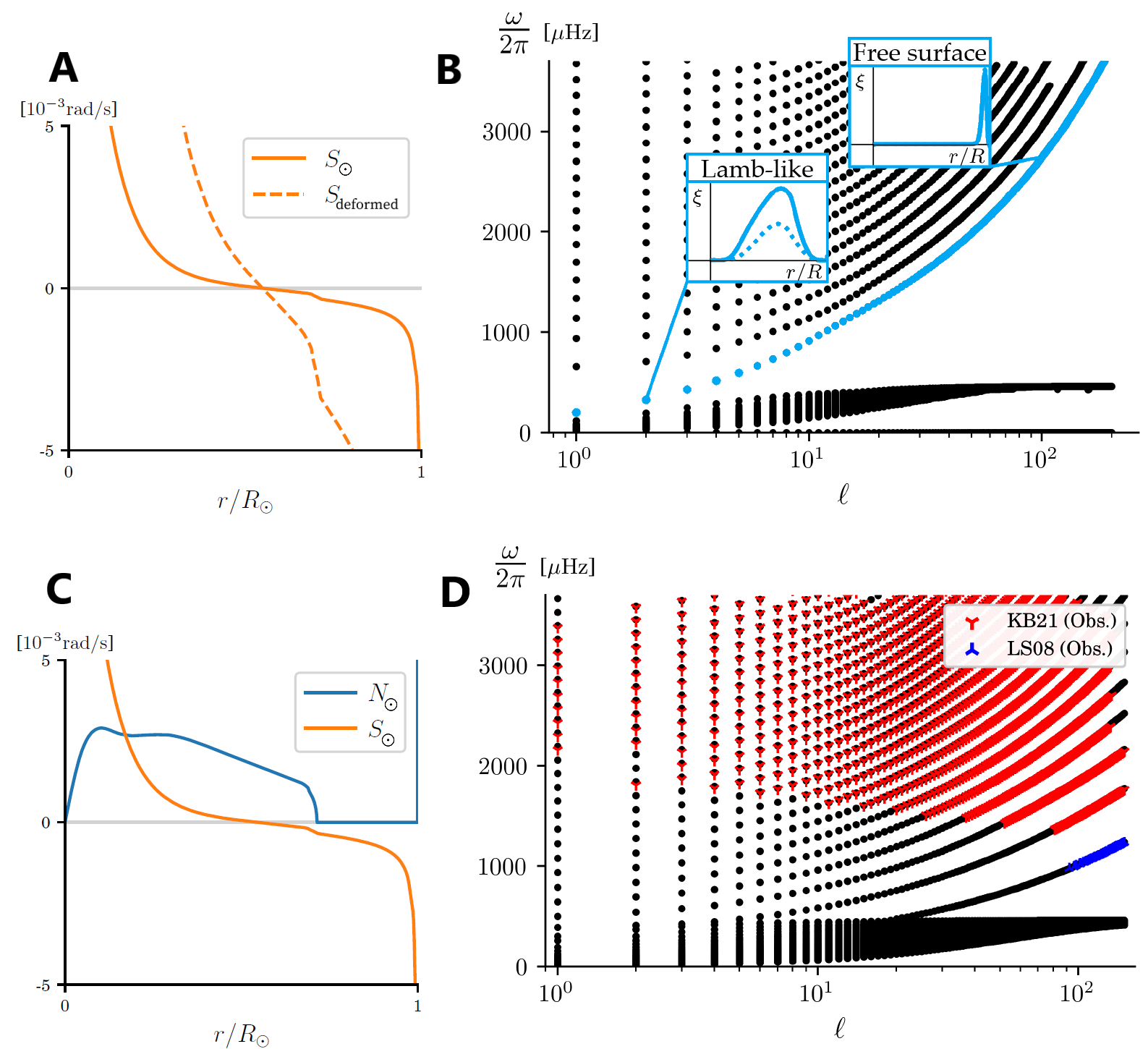}
    \caption{\textbf{Normal linear adiabatic modes of oscillations.} \textbf{(A)} Model and \textbf{(B)} frequencies of a fictitious star with $S_\mathrm{deformed}(r) = 10S_{\odot}(r)$. The spectrum show that the low-$\ell$ part of the $f$ branch consists in bulk modes localized where $S = 0$. The radial $\xi_r$ and horizontal $\xi_\mathrm{h}$ displacements are represented by dotted and solid lines, respectively. \textbf{(C)} Model and \textbf{(D)} frequencies of the modes of our solar model for the first hundreds of harmonic degree $\ell$. Linear theory predicts frequencies matching observed modes (colored crosses indicate data from \cite{larson2008,kiefer2021}). Lamb-like modes are difficult to identify when the $f$-branch penetrates the $g$-band.}
    \label{fig:S-fmode-solarSpectrum}
\end{figure*}

To show clearly these two regimes of $f$-modes, we create a model of a fictitious star, in which the buoyant-acoustic frequency $S_{\rm deformed}$ has been deformed from the solar case $S_{\odot}$, as presented in the panel A of Fig.~\ref{fig:S-fmode-solarSpectrum}. We then numerically solve for the frequencies of linear eigenmodes of this model (see Appendix \ref{apdx:linear_theory}). Looking at the wave spectrum in the panel B of Fig. \ref{fig:S-fmode-solarSpectrum}, we can see  that for $\ell \lesssim 15$ the $n=0$ mode propagates in the bulk of the star as a Lamb-like wave close to the $S_{\rm deformed} = 0$ radius, while for $\ell \gtrsim 15$, it is confined at the surface. 
In the case of the Sun, the repulsive effect of $L_\ell$ overcomes the trapping at $S_{\odot}=0$, causing Lamb-like modes at low $\ell$ to delocalize, making it more difficult to clearly distinguish the two regimes. This situation is atypical in topological physics and explains why solar Lamb-like waves are difficult to identify at first glance (see Fig.~\ref{fig:boosting} of Supplementary Materials).

\section{Non-linear compressible 2D simulations with MUSIC}

In this study, we demonstrate the natural excitation of Lamb-like waves in the Sun. Similar to acoustic waves, Lamb-like waves can only propagate in compressible media, necessitating the use of a fully compressible hydrodynamics code. Given that both anelastic and Boussinesq approximations inherently filter out acoustic waves including Lamb-like, these methods are unsuitable for the present investigation.

We present a 2D simulation of the solar interior using the \textsc{music} code \citep{Viallet13,Viallet16, Goffrey17}, one of the few stellar hydrodynamic codes that solves the fully compressible equations of hydrodynamics. This code uses a time-implicit method to solve the continuity, the momentum and internal energy equations for a fully compressible, inviscid fluid, which are described in Appendix \ref{apdx:music}. The initial conditions of the 2D \textsc{music} simulation are obtained from a 1D structure of the Sun computed with the stellar evolution code \textsc{mesa} \citep{Paxton2011}, calibrated to match observed characteristics of the Sun (see Appendix \ref{apdx:1Dmodel}). We will refer to the latter as the 1D model in the following.
Linear modes of this 1D model agree excellently with the frequencies of the observed solar oscillation modes, as shown in panel D of Fig.~\ref{fig:S-fmode-solarSpectrum}, confirming that it accurately describes the Sun. 

The numerical domain of the \textsc{music} simulation is a two-dimensional shell shown in the left panel of Fig.~\ref{fig:results-sim}, covered by two spherical coordinates: the radius $r$ and the co-latitude $\theta$. This domain is large enough to model realistic properties of waves and their excitation by convective motions \citep{LeSaux2022}. 
The boundary conditions are described in details in Appendix \ref{apdx:2Dsims}. In order to conclusively prove that Lamb-like waves propagate in the Sun, we apply solid boundary conditions at the outer surface for the radial velocity. This setup thus prevents the propagation of free surface gravity waves, ensuring that any $n=0$ mode consists exclusively of a Lamb-like mode. 

\section{Solar Lamb-like Waves Unveiled in Simulations}

\begin{figure*}
    %
    %\centering\includegraphics[width=\textwidth]{figures/fig2_V1_HSEsim_Vh_zoomSpectra.pdf}
    \plotone{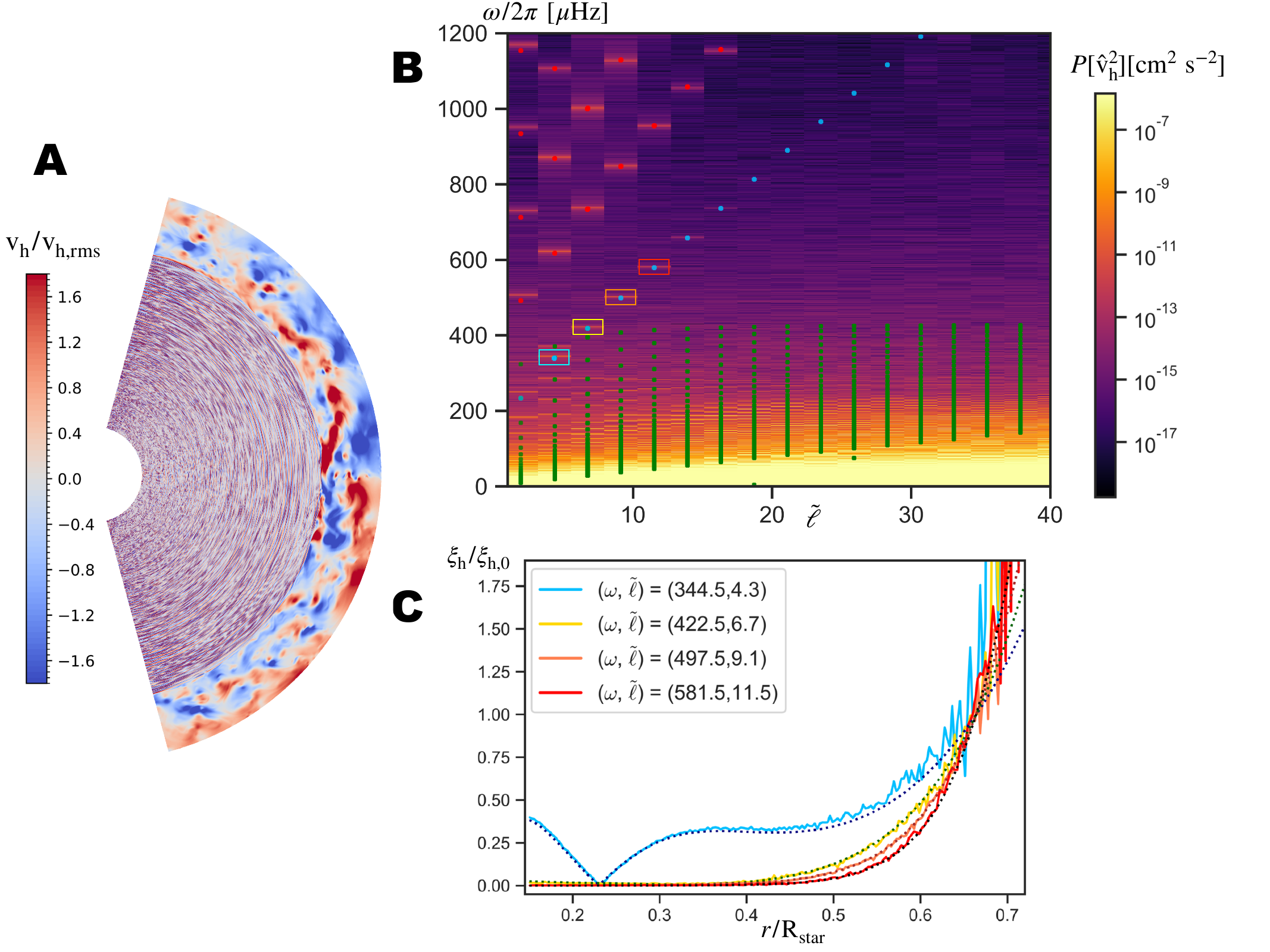}
    \caption{\textbf{Non-linear simulation of the solar interior shows the Lamb-like wave among the standing waves.} 
    The simulation results demonstrate that $f$-modes propagate within the Sun, driven by convective excitation, even at low values of $\ell$. \textbf{(A)} A snapshot of horizontal velocities, with values normalized at each radius by their root-mean-square. \textbf{(B)} The power spectrum of the horizontal velocity in the radiative zone displays peaks corresponding to normal modes. Red, blue, and green symbols indicate the numerically computed frequencies for linear standing $p$-modes, $f$-modes, and $g$-modes, respectively. The frequencies of the normal modes show excellent agreement with linear theory. At the surface and in observations, the power spectrum is more noisy due to convection and granulation. \textbf{(C)} When a normal mode frequency is selected, the radial profile of the horizontal displacement from the simulation (solid line) aligns perfectly with the numerically computed linear eigenfunction of the corresponding mode (dotted line). Displayed here are the profiles for a mixed $f$/$g$-mode (blue) and three pure $f$-modes (yellow, orange, red). The corresponding modes are marked by matching-colored boxes in the power spectrum in panel B.
    }
    \label{fig:results-sim}
\end{figure*}

Panel (A) of Fig.~\ref{fig:results-sim} shows a snapshot of the horizontal velocity in the simulation. The large-scale coherent flows, characteristic of convective motions in the Sun's envelope, appear prominently. In the radiative interior ($r<0.7\:R$), the tightly wound spirals are characteristic of the wavefronts of internal gravity waves. These features are similar to what is usually observed in solar hydrodynamical simulations in two- \citep{Rogers2006} or three-dimensions \citep{Brun2011, Alvan2014}.

The wave analysis is based on the methodology described in \citet{LeSaux2022, LeSaux2023}. To compute the power spectra of the velocity components $\vel_i$, with $i$ = $r$ or $\rm h$,  we first perform a temporal Fourier transform to obtain a pulsation $\omega$ dependence, and then use wedge harmonics to get an angular wavenumber $\elleff$ dependence, transforming $\vel_i(r,\theta,t)$ into $\hat{\vel}_i(r,\elleff,\omega)$. Wedge harmonics extend spherical harmonics to wedge-shaped domains, resulting in non-integer harmonic degrees $\elleff$ (see Appendix \ref{apdx:WH}). Panel (B) of Fig.~\ref{fig:results-sim} shows the power spectrum of the horizontal velocity $\velh$ as a function of frequency $\omega/2\pi$ and effective angular degree $\elleff$ averaged over a few radial cells around $r = 0.55 R_{\rm star}$. This spectrum is characteristic of the radiative zone. At the surface ($r = R_{\rm star}$), the frequencies of the modes would remain unchanged compared to the radiative zone, but the spectrum would be much more noisy due to surface convection.
Detection and identification of normal modes in the simulation is performed by looking for peaks in the amplitude of the power spectrum localized in ($\omega,\elleff$).
The colored symbols (red for $p$-modes, green for $g$-modes and blue for $f$-modes) are linear modes calculated independently. We observe excellent agreement between these frequencies and those measured in the simulation.
Taking a slice of the power spectrum at a specific $(\omega,\elleff)$, we extract the targeted mode, which provides the radial profile of the squared horizontal velocity $\vert\vel_h \vert^2(r)$ obtained in the simulation. This measurement is shown for four selected modes on panel (C) of Fig.~\ref{fig:results-sim}. The modes' amplitude are normalised to 1 at $r = 0.6 R_{\rm star}$ for better comparison (this radius was chosen to be close to the radiative boundary, but not too close to avoid contamination by convective boundary mixing). The modes ${(\omega / 2\pi,\elleff) = (428\,\mu {\rm Hz}, 6.7), (508\,\mu {\rm Hz}, 9.1) \;{\rm and} \;(588\,\mu {\rm Hz}, 11.5)}$, represented by the yellow, orange and red curves respectively, are $f$-modes. As predicted by Eq. \eqref{eq:schro}, the $f$-modes are confined to outer layers as $\elleff$, or equivalently $\ell$, increases. 
We compare the simulated velocities with the eigenfunctions of the corresponding linear modes (shown as dotted lines) and find excellent agreement. This demonstrates that Lamb-like waves are indeed excited and propagate within the Sun,  exhibiting the properties predicted by linear theory. Now, these Lamb-like waves reveal another notable feature: they can couple with other modes deep within the solar interior.

\section{Mixed $f/g$-modes in the Sun}

\begin{figure*}
    \centering
    \plotone{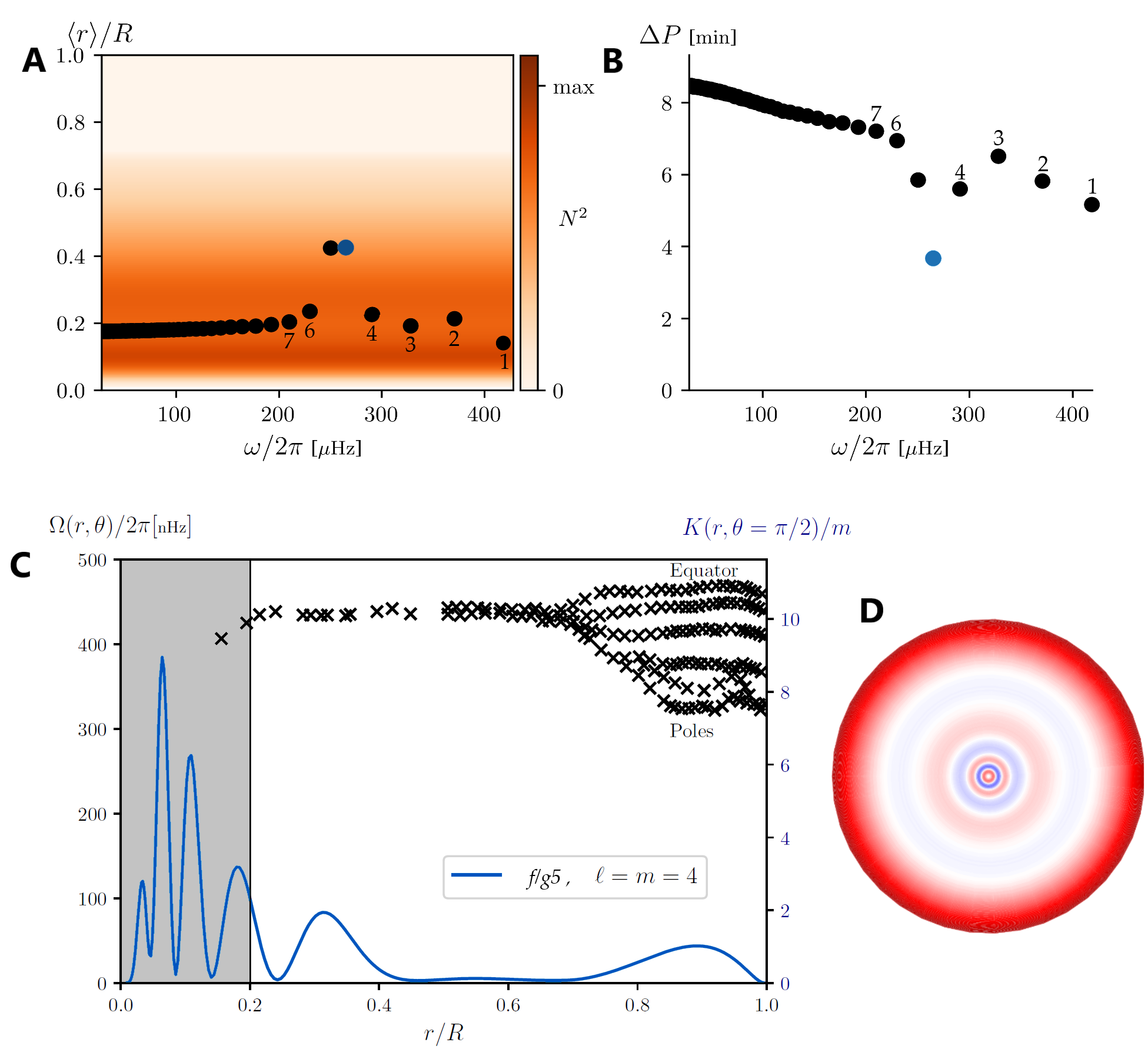}
    \caption{\textbf{Identification of mixed $f/g$-modes in the solar spectrum.} The $\ell = 4$ Lamb-like wave couples with the $g5$ mode to form a mixed mode. \textbf{(A)} The average position $\langle r \rangle$ of the $g$-modes defined in the main text is close to the maximum of buoyancy frequency $N^2$, except for two: they result from the coupling of the $f$-mode and the $g5$ mode, which thus have more amplitude in the outer layers. \textbf{(B)} The period spacing $\Delta P = P_{n+1}-P_n$ shows a dip at the fifth $g$-mode. \textbf{(C)} The kernel $K$ of the $f/g5$ mixed mode, in dark blue, shows significant amplitude in the solar core ($r/R_\odot<0.2$), implying a significant sensitivity to the core rotation rate. Dark crosses are data points for $\Omega$ from \cite{thompson2003} obtained observationally. \textbf{(D)} Representation of the horizontal displacement $\xi_\mathrm{h}$ of the mixed mode at frequency $265$ $\mu$Hz in the Sun (red for positive, blue for negative, linear scale). It has amplitude at the surface, as well as in the deep interior.}
    \label{fig:mixed_mode}
\end{figure*}

In the simulation, the mode at $(\omega/ 2\pi,\elleff)$ = (344 $\mu$Hz, 4.3) represented by the blue curve, shows a different structure than the other three $f$-modes: it has a node in the radiative zone. We identify here a mixed mode, resulting from the coupling of the $f$-mode with the $g_1$-mode. Mixed modes occur when two oscillations of different nature propagating in distinct layers of a star become coupled. One of the major successes of asteroseismology has been to predict and detect mixed $p$/$g$ modes in stars evolving beyond the main sequence \citep{scuflaire1974, Kjeldsen1995}, thus having amplitude not only in the deep core but also at the surface, making it possible to probe the entire interior of a star \citep[see][and references therein]{Hekker2017}. \\
The results of our simulation predict that mixed $f$/$g$ modes should exist in the Sun, at possibly different frequency and degree because of the shorter size of the simulated domain. By examining the linear modes of the full Sun (Fig.\ref{fig:S-fmode-solarSpectrum} D), we do find a mixed mode at $\ell = 4$ between the $f$-mode and the fifth $g$-mode oscillating at a frequency ${\omega}/{2\pi} = 265\;\mu$Hz, an oscillation period of almost an hour. We predicted the overlapping of $f$-modes and $g$-modes eigenfunctions from the properties of Lamb-like waves inherited from their topological origin. Indeed, for decreasing $\ell$, the $f$-mode frequency crosses the values of $g$-modes frequencies (Fig.\ref{fig:S-fmode-solarSpectrum} D), and has more and more amplitude in the interior (Figs.\ref{fig:S-fmode-solarSpectrum} B and Fig.\ref{fig:results-sim} C) leading to an overlap of their eigenfunctions. These two conditions cause the hybridization of the modes. \\
We show this hybridization by considering the kinetic energy-weighted average position of the modes
\begin{equation}
    \langle r \rangle = \frac{\int  \mathrm{d}r\; r \; r^2 \rho_0 (\xi_r^2 + \xi_h^2)}{\int \mathrm{d}r \; r^2 \rho_0 (\xi_r^2 + \xi_h^2)},
\end{equation}
which provides a diagnosis for mode coupling in the simulations. Panel (A) of Fig.~\ref{fig:mixed_mode} shows that the average position of the fifth $g$-mode is much higher than that of the other $g$-modes, all located close to the maximum of buoyancy frequency $N^2$ as expected for pure gravity modes. Panel (B) displays the standard diagnostic of mixed modes commonly used in observations \citep{ouazzani2020}. The periods $P_n$ of pure $g$-modes are expected to be nearly uniformly spaced \citep{unno1979,gough1993}. However, due to mode mixing, the period spacing $\Delta P \equiv P_{n+1}-P_n$ dips significantly at $n = 5$.\\
The value of the order $n$ of the $g$-mode implicated in the coupling with the $f$-mode is sensitive to the exact frequency of the latter, and could thus depend on the 1D model. Still, we find that a mixed $f/g$ mode at $\ell = 4$ happens with radial order $n=5$ also for the standard model S of the Sun \citep{christensen1996} and the seismic solar model of \citep{buldgen2020}, such that we do not expect a very different order in the Sun.

Mixed $f$/$g$ modes have not been noticed in the solar spectrum before. To our knowledge, the existence of a coupling between solar $f$ and $g$-modes was only studied in the dipolar case  $\ell=1$ when discussing the effect of Cowling's approximation \citep{christensen2001}.   \citet{provost2000} observed that the fifth solar $g$-mode at low $\ell$ exhibited a mixed nature, but did not realize this was due to its coupling with a deeply propagating $f$-mode. \citet{belkacem2022} noticed a coupling of $g$-modes with an external mode in recent seismic solar models as well, which in the light of our results appears to be these mixed $f$/$g$ modes.

\section{Sensitivity to the solar core rotation rate}

To date, the solar rotation rate $\Omega(r,\theta)$ has been accurately determined using $p$-modes in both the radiative and convective zones, but only for radii greater than 20\% of the solar radius. 
We demonstrate here that the mixed $f$/$g$5 mode identified above is very sensitive to the rotation rate of the solar core and thus provides a new method to potentially probe the innermost layer of the Sun. \\
Rotation causes a peak in the power spectrum of a given $(n,\ell)$ mode to split into $2\ell+1$ distinct peaks that correspond to each value of the azimuthal order $m$ upon action of the Doppler effect and Coriolis force. By measuring this splitting, one can infer the solar rotation rate $\Omega(r,\theta)$ by inversion \citep{Howe2009}. 
The split is generically expressed ${
    \Delta\omega_{n,\ell,m} \equiv \omega_{n,\ell,m}-\omega_{n,\ell,0} = \int K_{n,\ell,m}(r,\theta)\Omega(r,\theta)\mathrm{d}r\mathrm{d}\theta
}$, where $K_{n,\ell,m}$ is the kernel of the mode computed without rotation.
The expression of $K_{n,\ell,m}$ for a given mode is given in \citet{schou1994}. In the following, subscripts $(n,\ell,m)$ are omitted. 

To quantify how effectively a mode can probe the core rotation rate $\Omega^\mathrm{c}$, we define the sensitivity for solid body rotation of the core as
\begin{equation}
    s \equiv \frac{\partial \omega}{\partial \Omega^\mathrm{c}} = \int_{r<0.2 R_\odot} K \;\mathrm{d}r\mathrm{d}\theta,
\end{equation}
such that the splitting specifically due to the rotation of the core is $\Delta\omega_\mathrm{core} = s\,\Omega^\mathrm{c}$.
The solar modes used for inversion of rotation so far are acoustic and have limited sensitivity to core rotation since they do not travel in such deep layers. For instance, the mode $n=5$ and $\ell = m =1$, one of the deepest propagating $p$-modes, has $s = 7.27\times10^{-2}$. This implies that a core rotation of $\sim 400$ nHz would cause its splitting to increase by around 30 nHz. This change is too small to be conclusively measured with the current frequency resolution of approximately 10 nHz provided by the BiSON dataset \citep{howe2023}. 

In contrast, we find that the $f/g5$ mixed mode at $\ell=m=4$ has a sensitivity $s = 2.24$, thirty times more than $p$-modes. Figure~\ref{fig:mixed_mode} shows the substantial amplitude of the kernel $K$ of this mixed mode in the core, explaining its high sensitivity. For this mode, the known rotation rate in the radiative and convective regions account for a splitting of $786\;\mathrm{nHz}$, %
leading to the expression
\begin{equation}
    \Delta\omega/2\pi = 786\;\mathrm{nHz} + 2.24\;\Omega^\mathrm{c}/2\pi.
\end{equation}
The frequency resolution of 10 nHz in observations implies that the method presented here allows for the determination of $\Omega^\mathrm{c}$ with a precision of $\sim$5 nHz. For a hypothetical rotation rate of $\sim 400$ nHz, similar to that of the radiative region, the additional splitting would be around $10^{3}$ nHz. As a comparison, at $\ell=m=4$, a $g$-mode at $\omega/2\pi = 103\mu$Hz has $s = 3.16$. It is thus as sensitive as the mixed mode, but we argue that it is a more difficult observational target because of its evanescence in the radiative zone and its lower frequency.

From an observational perspective, the mixed $f$/$g$ mode is found in a frequency range that is not accessible yet with currently available data. However, using the \textsc{music} simulation, the eigenfunction computed from the 1D model and observational data from \citet{Davies2014}, we can estimate the relative surface amplitudes between different modes. We find that the mixed $f/g$-mode at 265 $\mu$Hz have a surface amplitude comparable to those of $p$-modes within the same frequency range, whereas $g$-modes in this range exhibit significantly lower surface amplitudes (see details in Appendix \ref{apdx:surf_amplitude}). Instead, we find that $g$-modes have comparable surface amplitudes at lower frequencies ($\sim$ 50–100 $\mu$Hz), where they are expected to have largest surface amplitudes \citep{belkacem2022}.
But as background noise from granulation and turbulence increases at lower frequencies in the solar spectrum, the detection threshold at 265 $\mu$Hz is two to three times lower \citep{belkacem2022}. Consequently, despite having similar sensitivities, we conclude that the higher frequency of the mixed $f/g5$ mode makes it an easier target and a more promising opportunity for probing the solar core in the future than pure $g$-modes.

\section{Summary \& Discussion}

For decades, $f$-modes have been thought to ``provide no information about the solar interior'' \citep{aerts2010}. Indeed, $f$-modes are often considered to be only surface gravity waves. We overturn this longstanding belief by demonstrating that low degree $f$-modes are bulk modes in the form of topological waves, in the process making a novel use of a topological wave as a probe rather than for transport properties as done in condensed matter physics. By performing a state-of-the-art simulation of the solar interior using the fully compressible hydrodynamical code \textsc{music}, we demonstrate that not only low-degree $f$-modes in the Sun have this topological origin, but also that they are naturally excited by solar convection. In addition, we identify a previously unnoticed mixed $f$/$g$ mixed mode of the Sun. The coupling is due to properties inherited from the topological origin of the $f$-mode at low degree which imposes that it propagates deep in the solar interior. This mixed mode, oscillating every hour, is found to exhibit significant amplitude both in the Sun's core and at its surface. Its sensitivity to the core’s rotation rate is over 30 times greater than currently observed modes. Thus, it provides a novel way to measure this rotation rate, that represents to date our most promising opportunity for future measurements as it is located in a less noisy region of power spectra than $g$-modes with expected maximum amplitudes (100 $\mu$Hz).\\
The argument is general, implying that topological mixed modes should universally be found in bodies having a stably stratified interior. We suspect that the $f$/$g$ mode found in Saturn is such a mode \citep{dewberry2021}. Our study highlights the key role of compressible hydrodynamics in modelling stellar interiors, presenting the first-ever simulation of the coupling between compressible and gravity modes. Altogether, this holds great promise for asteroseismology with the upcoming launch of the PLATO mission \citep{Rauer2014} in 2026, which will provide low degree oscillations of solar-like stars, whose cores could be accessed with mixed $f$/$g$ modes.

%\newpage

%% Please use the acknowledgment and contribution environments. This will 
%% be anonomyized when the "anonymous" style option is used. 
\begin{acknowledgments}
The authors thank T. Guillet and A. Morison for their support with \textsc{MUSIC}, as well as R. A. Garcia and S. Korzennik for fruitful discussions on observational perspectives. We gratefully acknowledge support from the PSMN (Pôle Scientifique de Modélisation Numérique) of the ENS de Lyon and the Alfven facility of CEA Paris-Saclay for the computing resources. 
A.L. and G.L. acknowledge funding from the European Research Council (ERC) CoG project PODCAST No. 864965. A.L. is funded by Contrat Doctoral Spécifique Normalien. A.L.S. acknowledges support from the European Research Council (ERC) under the Horizon Europe programme (Synergy Grant agreement 101071505: 4D-STAR). While partially funded by the European Union, views and opinions expressed are however those of the author only and do not necessarily reflect those of the European Union or the European Research Council. Neither the European Union nor the granting authority can be held responsible for them.
\end{acknowledgments}

\begin{contribution}
%%This section gives authors the space to recognize author contributions. The text inside this environment is NOT counted towards the total word quanta. At a minimum, manuscripts are expected to include this text:

A.L.S. ran the simulations, calibrated the 1D model, and performed their post-processing analysis. A.L. computed the linear spectra and the inversion kernels. All authors participated in the redaction of the manuscript. Numerical scripts and data are available on a Zenodo repository at
\href{doi.org/10.5281/zenodo.15600943}{doi.org/10.5281/zenodo.15600943}.

%% But authors are expected to provide more specific details, e.g. 
%%
%%SC was responsible for writing and submitting the manuscript.
%%WWM came up with the initial research concept and edited the manuscript.
%%OTS obtained the funding and edited the manuscript.
%%EBF provided the formal analysis and validation. He also edited the manuscript.
%%GEH Supervised the undergraduates, wrote the software and administers the project github and Zenodo repositories.
%%
%% Authors can use the Contributor Role Taxonomy (CRediT) at
%% for ideas on how write a good statement tailored to their needs.

\end{contribution}

%\facilities{HST(STIS), Swift(XRT and UVOT), AAVSO, CTIO:1.3m, CTIO:1.5m, CXO}

%%          Cloudy \citep{2013RMxAA..49..137F}, 
%          Source Extractor \citep{1996A&AS..117..393B}
%          }


\appendix

\section{Linear theory}
\label{apdx:linear_theory}
The Sun is stratified by gravity and maintains a state of hydrostatic equilibrium. Two types of motion perturb this static balance. In the convective zone, turbulent motion occurs in the form of convective eddies driven by the convective instability. In addition, acoustic and internal gravity waves propagate inside different layers of the Sun, depending on their nature and frequency \citep{cowling1941,unno1979,gough1993,aerts2010}. These waves are primarily excited by the convective motion and are classified by their normal modes, with frequencies characteristic of the Sun. Since the background structure through which they propagate is spherically symmetric, the normal modes generally take the form $\mathrm{e}^{-i\omega t}Y_\ell^m(\theta,\phi)f(r)$, where $\omega$ is the frequency, $Y_\ell^m(\theta,\phi)$ is a spherical harmonics. $f$ is a function to be determined for each component of the wave (radial and horizontal velocities, pressure, density, temperature).

Assuming linear and adiabatic perturbations, along with no perturbations in the gravitational potential \citep[Cowling's approximation][]{cowling1941}, the normal modes of the sun are solutions of
\begin{equation}
    \omega X=\mathcal{H} X,\quad\mathrm{with}\quad
    \mathcal{H}\equiv
    \begin{pmatrix}  0 & 0 & 0 & L_\ell(r)\\
    0 & 0 & iN & -iS +\frac{i}{2} c_{\rm s}'+ ic_{\mathrm s}\partial_r\\
     0 & - iN  & 0 & 0\\
    L_\ell(r) & iS +\frac{i}{2} c_{\rm s}'+i c_{\mathrm s}\partial_r & 0 & 0\\ \end{pmatrix},\label{eq:linear_modes}
\end{equation}
where the perturbation vector $X(r) \equiv  \begin{pmatrix} \Tilde{v_\mathrm{h}} & \Tilde{v_r} & \Tilde{\Theta} & \Tilde{p} \end{pmatrix}^\top$ is based on re-scaled perturbed quantities of the horizontal velocity, radial velocity, entropy and pressure respectively, by

\begin{equation}
  \begin{aligned}
    v' &\mapsto  \Tilde{{v}} = \rho_0^{1/2}r \;v' = \Tilde{v_r}(r){Y}_\ell^m \mathrm{e}_r + \Tilde{v_\mathrm{h}}(r)\frac{ir}{\sqrt{\ell(\ell+1)}} {\nabla}(Y_\ell^m \mathrm{e}_r) ,\\       
    p' &\mapsto \Tilde{p} = \rho_0^{-1/2}c_\mathrm{s}^{-1}r\;p',\\
\rho' &\mapsto \Tilde{\Theta} = \rho_0^{-1/2}r\frac{g}{N}\;(\rho' - \frac{1}{c_\mathrm{s}^2}p'),
  \label{eq:transform}
  \end{aligned}
\end{equation}
The buoyant-acoustic frequency $S$ is given by Eq.~\eqref{eq:stellarS}, the buoyancy frequency squared is $N^2 \equiv -g\dr{\ln\rho_0} - \frac{g^2}{c_\mathrm{s}^2}$, the Lamb frequency squared ${L_\ell^2 \equiv c_\mathrm{s}^2 \; \ell(\ell+1)/r^2}$ and $c_\mathrm{s}' \equiv \dr{c_\mathrm{s}}$. Equation~\eqref{eq:linear_modes} must be accompanied by appropriate boundary conditions. Solar boundaries require $v_r = 0$ at $r=0$ and a free surface condition at the surface $\partial_t p + v_r\dr{P_0} = 0$. The numerical wedge domain used in \textsc{music} requires $v_r = 0$ at $r = r_\mathrm{out}$.

\begin{figure}
    \centering
    \plotone{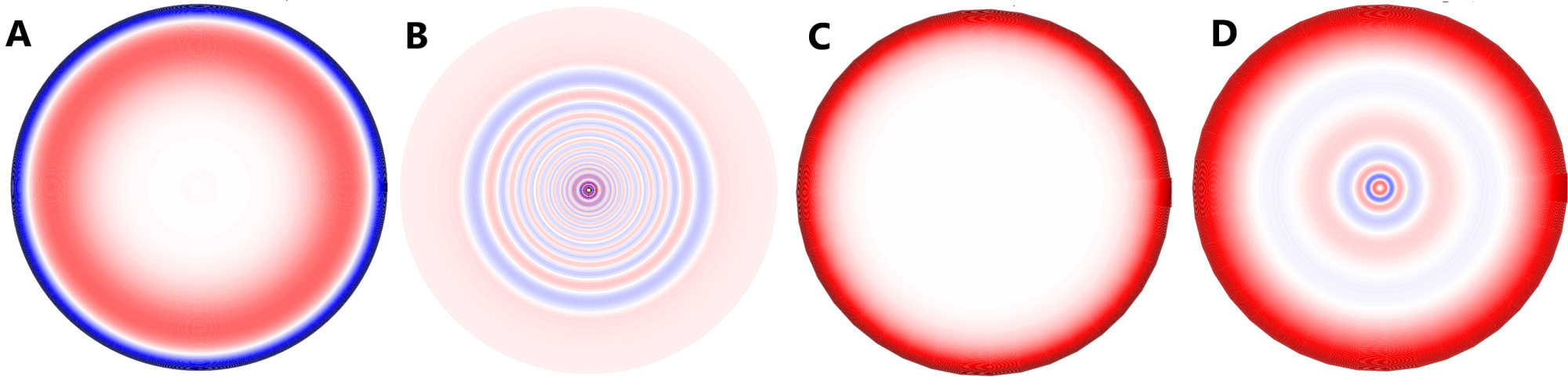}
    \caption{\textbf{Modes of different types are located in different regions of the Sun.} Red and blue are regions of positive and negative horizontal displacement $\xi_\mathrm{h}$. \textbf{(A)} is the $n=1$, $\ell=4$ $p$-mode, \textbf{(B)} is the $n=35$, $\ell=4$ $g$-mode, \textbf{(C)} is the $\ell=10$ $f$-mode, \textbf{(D)} is the mixed $f/g5$ mode discussed in the main text.}
    \label{fig:eigenfunctions_pgfmm}
\end{figure}

The eigenfrequencies and eigenvectors of $\mathcal{H}$ are obtained by solving Eq.~\eqref{eq:linear_modes} numerically, for a given model of the Sun which provides $N(r)$, $S(r)$ and $c_\mathrm{s}(r)$. The numerical procedure utilizes the EVP class of the \textsc{dedalus} code \citep{burns2020}. It allows us to find the set of $(\omega,X(r))$ which satisfy Eq.~\eqref{eq:linear_modes} for any value of $\ell$. \textsc{Dedalus} uses spectral methods which decompose solutions on $N_r$ Chebyshev polynomials, in order to obtain a matrix eigenvalue problem which is then solved by linear algebra techniques. This method can capture eigenmodes with radial order up to $N_r$. Examples of eigenfunctions of different solar modes are shown in Fig.~\ref{fig:eigenfunctions_pgfmm}.

\section{Solar 1D model}
\label{apdx:1Dmodel}
To investigate the oscillation modes of the Sun, we construct a one-dimensional model of the Sun using \textsc{mesa} \citep{Paxton2011, Paxton2013, Paxton2015, Paxton2018, Paxton2019, Jermyn2023}. This involves calibrating a 1D solar model with the \textit{simplex solar calibration} routine provided by \textsc{mesa}. This method aims at minimizing a $\chi^2$ function, which is defined as
\begin{equation}
    \chi^2 = \sum_j \frac{(O_j - M_j)^2}{\sigma_j^2},
    \label{eq:chi2}
\end{equation}
with $O_j$ the observed characteristics of the Sun and $M_j$ the modelled ones. These characteristics used to calibrate the model are the luminosity L, the radius R, surface metals $\rm (Z/X)_{\rm surf}$ and helium $\rm Y_{\rm surf}$ abundances, the position of the interface between the radiative and the convective zones $\Rcz$ and the sound speed profile of the solar interior. The measured values of L and $\teff$ are the ones given by the \textit{Resolution B3} of the International Astronomical Union \citep{Mamajek2015} and are the default values used in \textsc{mesa}. For the metals surface abundances, we use the values recently determined by \citet{Magg2022}, which are similar to those from \citet{Grevesse1998} and correspond to the high metal abundance case.  The helium surface abundance is based on measurements by \citet{Basu2004}, and the position of the radiative/convective interface is taken from \citet{Basu1997}. For the sound speed profile, we use the default profile in \textsc{mesa}, which is derived from helioseismology by \citet{Basu1996}. All observed values for these parameters are summarized in Table  \ref{tab:charac_Sun}, with the associated uncertainties $\sigma_j$, except for the sound speed profile, for which we refer the reader to the original paper \citep{Basu1996}.

\begin{table}[t]
   \caption{\textbf{Characteristics of the Sun and of our solar model}. Obtained after calibration.}
   \label{tab:charac_Sun}
   \centering
    \begin{tabular}{c c c c c} 
     \hline \hline
     Parameter  &  Obs. & $\sigma_j$ & Model & Model/Obs \\
      \hline
      Age $(\rm Gyr$) & $4.57$ & - & $4.57$ & 1.0\\
      M ($\rm g$) & $1.9884 \times 10^{33}$ & - & $1.9884 \times 10^{33}$ & 1.0 \\
      $\teff$ (K) & $5772.0$ & - & $5760.1$ & 0.998 \\
      R ($\rm cm$) & $6.957 \times 10^{10}$ & $3.211 \times 10^{8}$  &$6.9575 \times 10^{10}$ & 1.00007 \\
      L ($\rm erg.s^{-1}$) & $3.828 \times 10^{33}$ & $1.4 \times 10^{30}$ & $3.797 \times 10^{33}$ & 0.992\\
      $\rm (Z/X)_{\rm surf}$ & $0.0252$ & $10^{-3}$ & $0.0259$ & 0.972 \\
      $\rm Y_{\rm surf}$ & $0.2485$ & 0.0035 & $0.2537$ & 1.021 \\
      $\Rcz$/R$_\odot$ & $0.713$ & $10^{-3}$ & 0.714 & 0.999 \\
      \hline
   \end{tabular}
\end{table} 

For the calibration process, the parameters to be adjusted are the iron-to-hydrogen number ratio [Fe/H] = $\log \left( (Z/X)/(Z/X)_{\odot} \right)$ and the helium abundance Y in the convective zone, as well as the mixing length $\alpha_{\rm MLT}$ and the overshooting $f_{\rm ov}$ parameters. The subscript in $(Z/X)_{\odot}$ denotes the reference solar value, and it is equal to the one given for surface value, $\rm (Z/X)_{\rm surf}$, in Table \ref{tab:charac_Sun}. In evolutionary 1D models, convective mixing is assumed to be instantaneous, thus the convective region is assumed homogenous. 
As a result of the calibration process, the obtained iron-to-hydrogen number ratio is [Fe/H] = 0.0531, the helium abundance is Y = 0.25375, the mixing length and overshooting parameters are $\alpha_{\rm MLT}$ = 2.0678 and $f_{\rm ov}$= 0.0253. For this best calibrated model, the value of the loss function is $\chi^2$ = 4.33. The results of the calibration are presented in Table \ref{tab:charac_Sun}, where the last column compares our solar model parameters to the observed values. The parameters presented in Table \ref{tab:charac_Sun} for which no uncertainties are given, were not included in the calibration process.

To validate the seismic accuracy of our calibrated model, we compute the eigenfrequencies of its oscillation modes using the \textsc{Dedalus} code as described above. The right panel of Fig. \ref{fig:S-fmode-solarSpectrum} confirms that our model accurately reproduces the eigenfrequencies for the global oscillation modes of the Sun, allowing for a precise comparison with helioseismic observations. In this figure, the red crosses represent the observed oscillations frequencies of $p$-modes measured by \citet{kiefer2021} and the blue crosses the oscillation frequencies of the $f$-modes measured by \citet{larson2008}. This good agreement between our model and the observed oscillations frequencies is the result of using the sound speed profile for the calibration. Indeed, the sound speed profile itself is inferred from helioseismology \citep{Basu1996}. The 1D model used in this study is publicly available, as well as the numerical treatment of the linear theory.

\section{Numerical simulations}
\subsection{The \textsc{music} code}
\label{apdx:music}
We present a 2D simulation of the solar interior using the stellar hydrodynamics code \textsc{music} \citep{Viallet16, Goffrey17}. This code employs a time-implicit method \citep{Viallet13} to solve for the mass, the momentum and internal energy for a fully compressible, inviscid fluid, i.e.,

\begin{equation}
\frac{\partial \rho}{\partial t} = - \vec{\nabla} \cdot (\rho \vec \velvec),
\end{equation}

\begin{equation}
\frac{\partial\rho \vec\velvec}{\partial t} = - \nabla \cdot (\rho \vec \velvec \otimes \vec \velvec) - \vec{\nabla}p + \rho\vec{g},
\end{equation}

\begin{equation}
\frac{\partial \rho e}{\partial t} = - \vec{\nabla} \cdot (\rho e \vec\velvec) - p \vec{\nabla} \cdot \vec \velvec - \vec{\nabla} \cdot \vec{F_r} + \rho \epsilon_{\rm nuc},
\end{equation}

where $\rho$ is the density, $e$ the specific internal energy, $\vec \velvec$ the velocity field, $p$ the gas pressure, $\epsilon_{\rm nuc}$ is the specific energy released by nuclear burning and $\vec{g}$ the gravitational acceleration. The radial profile of nuclear energy $\epsilon_{\rm nuc}$  is taken from the 1D model, and is assumed constant during the simulation time, as nuclear burning evolves on much longer timescales. The hydrodynamical simulation run for this work assumes spherically symmetric gravitational acceleration, $\vec{g} = - g \vec {\rm e}_r$, which is updated after each time step:
\begin{equation}
    g(r) = 4\pi \frac{G}{r^2} \int^r_0 \overline{\rho}(u)u^2 {\rm d}u,
\end{equation}
with $\overline{\rho}(r)$ radial density profile given by
\begin{equation}
    \overline{\rho}(r) = \AvgS{\rho}
\end{equation}
where the operator $\AvgS{.}$ is an angular average over the whole unit sphere, defined as
\begin{equation}
    \left< h \right>_{\mathcal{S}} \coloneqq \frac{1}{4\pi} \int_\mathcal{S} h(\theta, \phi) \, 2\pi \sin \theta {\rm d} \theta {\rm d} \phi.
    \label{eq:angular_av}
\end{equation}

For the solar simulations considered in this work, the major heat transport that contributes to thermal conductivity  is radiative transfer characterised by the radiative flux  $\vec{F_r}$, given within the diffusion approximation by

\begin{equation}
\vec{F_r} = - \frac{16 \sigma T^3}{3\kappa \rho} \vec{\nabla} T = - \chi \vec{\nabla} T,
\label{eq:radiative_flux}
\end{equation}
with $\kappa$ denotes the Rosseland mean opacity of the gas, $\sigma$ the Stefan–Boltzmann constant and $\chi$ the thermal conductivity respectively.
\textsc{music} incorporates realistic stellar opacities \citep[OPAL, ][]{Iglesias1996} for solar metallicity and equations of state \citep[OPAL EOS, ][]{Rogers2002} suitable for describing the solar interior.

\subsection{2D solar simulations}
\label{apdx:2Dsims}
The radial extent of the simulation described in the main text spans from $r_{\rm in} = 0.15 R_{\rm star}$ to $r_{\rm out} = 0.9 R_{\rm star}$. The co-latitudinal range is $\theta \in \left[\frac{\pi}{12};\frac{11\pi}{12}\right]$. We use a grid resolution of $N_r \times N_{\theta} = 1024 \times 1024$ cells, with uniform distribution in the radial direction. The size of a unit radial grid cell is $\mathrm{d}r = 5.1 \times 10^7$cm. The characteristic size of an angular grid cell is $\mathrm{d}\theta = 2.3 \times 10^{-3}$ rad. The resolution in the $\theta$-direction is defined by the requirement to preserve a satisfying aspect ratio of the grid cells on the whole domain on a spherical grid, ensuring adequate resolution for both the excited wave and convective motions \citep{baraffe2021, Vlaykov2022, LeSaux2022}. The simulation takes into account the variable helium mass fraction Y, which is set from the 1D model and do not evolve during the simulation, and we impose the metallicity Z = 0.02.

%tau_conv - 5.4 x 10^5 s -> omega_conv = 1.85 uHz

The boundary conditions in the radial direction are the imposition of a constant radial derivative on the density at the inner and outer radial boundaries \citep[see discussion in][]{Pratt2016}. The energy flux at both $\rin$ and $\rout$ is fixed and equal to the value of the energy flux at the corresponding radii in the calibrated 1D model. In terms of velocity, we impose reflective conditions at the radial boundaries as
\begin{itemize}
\item $\vel_r = 0$ and $\frac{\partial \vel_{\theta}}{\partial r} = 0$ at $\rin$ and $\rout$.
\end{itemize}

The boundaries in the latitudinal directions are periodic for the density, the energy and the velocities. For instance, for the velocity this is expressed as
\begin{itemize}
\item $\vel_r(\theta_{\rm min}) = \vel_r(\theta_{\rm max})$,
\item $\vel_{\theta}(\theta_{\rm min}) = \vel_{\theta}(\theta_{\rm max})$,
\end{itemize}
with $\theta_{\rm min} = \pi/12$ to $\theta_{\rm max} = 11\pi /12$ in this case.

\begin{figure}
    \centering
    \plotone{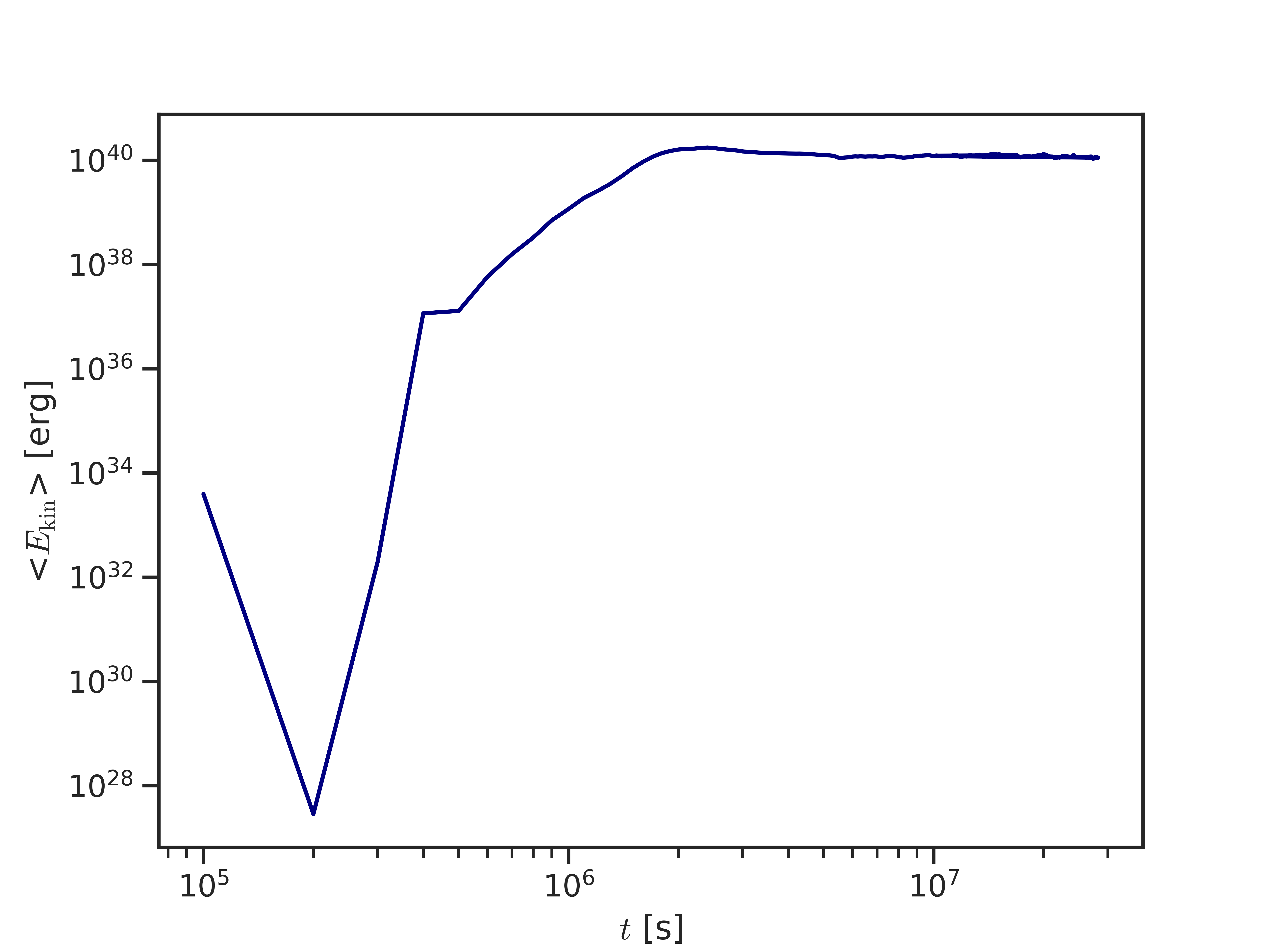}
    \caption{\textbf{Evolution of the total kinetic energy in the simulation over time.} The convective instability begins to develop around $2 \times 10^{5}$ s and then saturates as the convective layer fully develops. The simulation reaches a dynamical steady state once the plateau in kinetic energy is reached, around $3\times10^{6}$ s. The wave analysis is performed after that time.
    }
    \label{fig:EkvsTime}
\end{figure}

In panel A of Fig. \ref{fig:results-sim} shows a snapshot of the horizontal velocity in the simulation as a function of radius and co-latitude. For better visibility, we normalize its values at each radius by the root-mean-square value of the horizontal velocity. Indeed, the amplitude of the velocity in the radiative zone is much smaller than in the convective zone. The root-mean-square of the component $\vel_i$ of the velocity $\Virms$ is defined as

\begin{equation}
    \Virms(r) \coloneqq \sqrt{\AvgST{\vel_i^2(r,\theta, t)}},
    \label{eq:vrrms}
\end{equation}
with $i = r,$ h denotes the radial and horizontal components of the velocity vector. The operator $\AvgS{.}$ is defined in Eq. \ref{eq:angular_av} and $\AvgT{.}$ denotes the temporal average, and it is defined by
\begin{equation}
    \left< f(t) \right>_t \coloneqq \frac{t_0}{T} \int_0^T f(t) {\rm d} t,
     \label{eq:avgT}
\end{equation} 
with $t_0$ the time when the simulation reaches steady state convection and $T$ the final time of the simulation. To estimate when the convection reaches a steady state, we plot on Fig. \ref{fig:EkvsTime} the evolution of the total kinetic energy in the numerical model as a function of time. The initial peak in kinetic energy at comes from strong acoustic waves generated at the start of the simulation. Then, convection starts around $2 \times 10^5$ s and then reaches a plateau around $3 \times 10^6$ s, which define the value of $t_0$ in Eq. \eqref{eq:avgT}. We then run the simulation for a total time of $T \sim 3 \times 10^7$ s.

\section{Wedge harmonics decomposition}
\label{apdx:WH}
The wedge harmonics basis is a basis of eigenfunctions of the Laplacian on wedge-shaped domains that do not span an entire hemisphere in the latitudinal direction, as in the case used in this study \citep[see Apendix C of][]{LeSauxPhD}. It is a generalization of the spherical harmonics basis, designed for wedges. This alternative choice of angular basis is the reason why the power spectra computed from \textsc{music} work with effective angular degrees $\elleff$ that take non integer values, rather than the classical angular degree $\ell$ used in spherical harmonics. $\elleff$ still measures an angular wavenumber of waves with typical angular wavelength $\pi/{\elleff}$. In the case of the spherical harmonics, eigenvalues of $-r^2\Delta$ on the sphere are $\lambda_\mathrm{sphere} = \lambda_{\ell}^{\rm SH} = \ell (\ell + 1)$. By equating the eigenvalues in the spherical harmonics case and wedge harmonics cases, an effective angular degree $\elleff$ can be defined such that
\begin{equation}
    \lambda_{\rm wedge} = \lambda_{\elleff}^{\rm WH} = \elleff(\elleff+1) ,
\end{equation}
yielding
\begin{equation}
    \elleff = \sqrt{\lambda_{\elleff}^{\rm WH} + \frac{1}{4}} - \frac{1}{2}.
\end{equation}
The wedge harmonics basis and effective wavenumbers are only consequences of the removal of the north and south poles in the simulation, and have no impact on the underlying physics. For the wedge geometry used in the simulation, i.e. $\theta \in \left[\frac{\pi}{12};\frac{11\pi}{12}\right]$, the first values of $\elleff$ are \\${(0, 1.9, 4.3, 6.7, 9.1, 11.5, 13.9,...)}$.

\section{Effective potential theory}
\label{apdx:potential}
\begin{figure}
    \centering
    \plotone{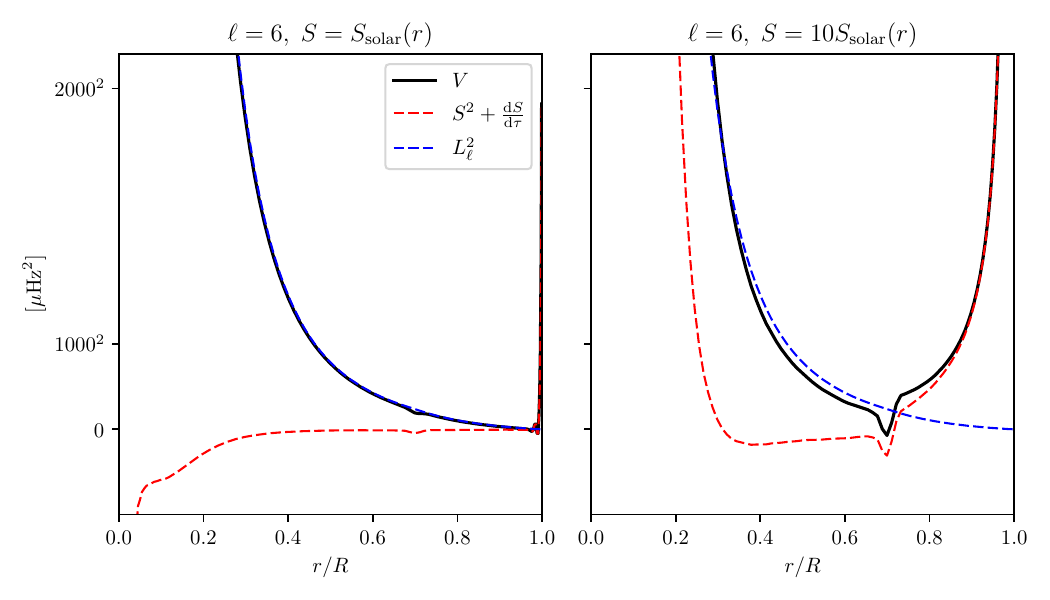}
    \caption{\textbf{Effective potential of Eq.~\ref{eq:schro} for the $p$- and $f$-modes in the high-frequency limit}. In he case $\ell = 6$, the minimum of the potential in the Sun is at the surface. In the fictitious star discussed in the main text for which $S(r) = 10 S_{\odot}(r)$, the minimum is located inside the bulk of the star, where the $f$-mode is much closely described by a Lamb-like mode.}
    \label{fig:potentials}
\end{figure}

To highlight why both surface-intensified waves and interior-localized waves can exist on the $f$-branch, we derive an effective potential theory within the Cowling approximation.
Building on the work of \citet{vorontsov1989}, we find that in the high-frequency limit $\omega \gg N$, the linear perturbation equations reduce to the Schrödinger equation
\begin{equation}
    \left(-\frac{\mathrm{d}^2}{\mathrm{d}\tau^2} + S^2 + \frac{\mathrm{d}S}{\mathrm{d}\tau} + L_\ell^2 \right)\Bar{p} = \omega^2\Bar{p},
    \label{eq:schro}
\end{equation}
with effective potential $V_{\rm eff} \equiv S^2 + \frac{\mathrm{d}S}{\mathrm{d}\tau} + L_\ell^2$. $\tau$ denotes the acoustic radius ($\mathrm{d}\tau = \mathrm{d}r/c_\mathrm{s}$), $\Bar{p} \equiv \frac{r}{\sqrt{\rho_0 c_\mathrm{s}}}p^\prime$ and $p^\prime$ is the Eulerian pressure perturbation and $L_\ell = c_\mathrm{s}\sqrt{\ell(\ell+1)}/r$ is the Lamb frequency measuring the angular wavelength in spherical geometry. The competition between $S^2 + \frac{\mathrm{d}S}{\mathrm{d}\tau}$ and $L_\ell^2$ in $V_{\rm eff}$ determines whether the modes are localized in the bulk of the star, where $S^2 + \frac{\mathrm{d}S}{\mathrm{d}\tau}$ is minimum, or in the outer layers where $L_\ell^2$ is minimum. Figure~\ref{fig:potentials} shows profiles of effective potential $V_{\rm eff}$ at $\ell = 6$ obtained for the Sun and a fictitious star, where the buoyant-acoustic frequency has been amplified $S= 10S_{\odot}$. 

For the fictitious star, the effective potential is dominated by the contribution of $S^2 + \frac{\mathrm{d}S}{\mathrm{d}\tau}$, as in the academic case derived by \citet{leclerc2022}. As such, topological waves are trapped locally in the bulk of the star. For the Sun, the contribution of $S^2 + \frac{\mathrm{d}S}{\mathrm{d}\tau}$ is counterbalanced by the repulsive effect of $L_\ell^2$, implying mode delocalization.

\section{Distorsion of the buoyant-acoustic frequency $S(r)$}
\label{apdx:distortS}
 In order to confirm that the $f$-modes at low $l$ are indeed of topological origin, we deform continuously deform the values of $S(r)$ in order to make the contribution $S^2 + \frac{\mathrm{d}S}{\mathrm{d}\tau}$ dominate the effective potential, similarly to the normal form analysed in \cite{leclerc2022}. This academic situation corresponds to a case of high coupling, where $S$ exhibits large positive and negative values around the cancellation point and varies rapidly. In this regime, the Lamb-like wave consists solely of horizontal velocity with no radial velocity, and its eigenfunctions are confined around the cancellation point of $S(r)$.
 
\begin{figure}
    \centering
    \plotone{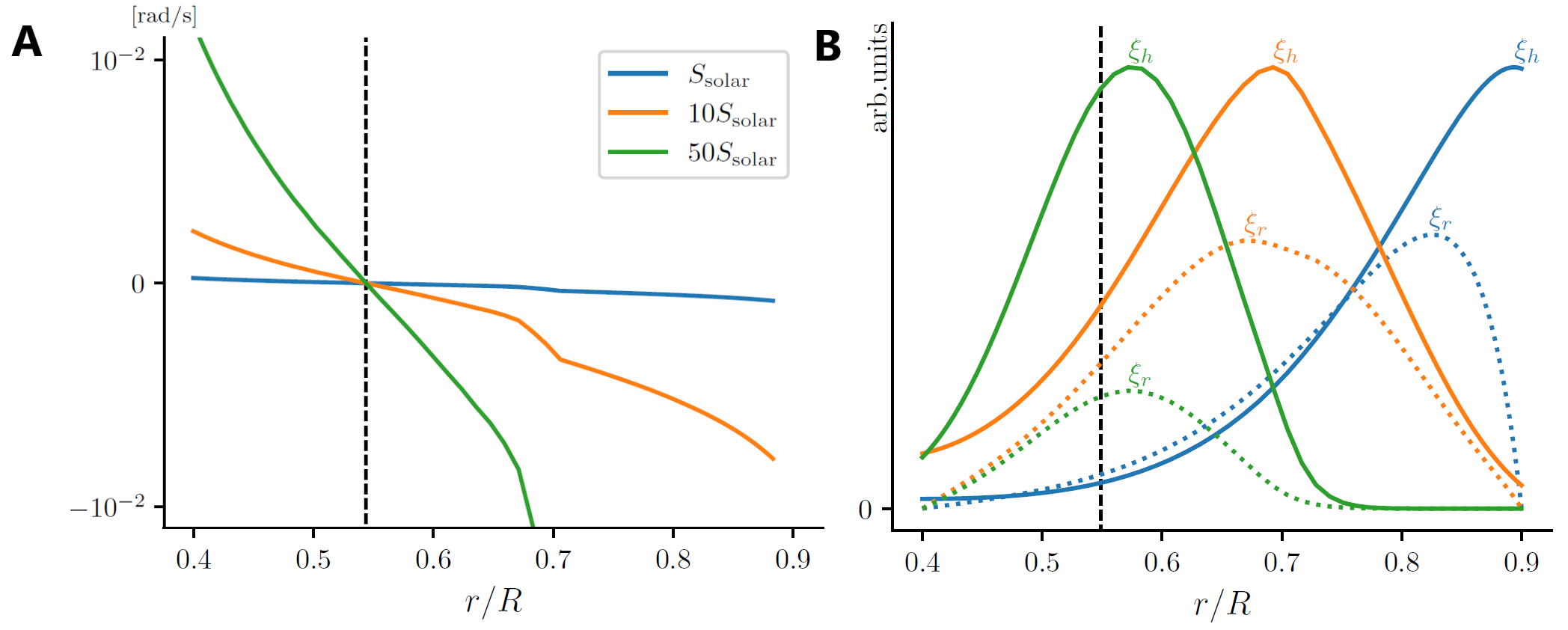}
    \caption{\textbf{Displacements of the Lamb-like wave for increasing values of $S(r)$.} \textbf{(A)} Three deformed profiles of $S(r)$. \textbf{(B)} Displacements of the $n=0$, $\ell = 4$ modes for the 3 profiles. As the values of $S$ increase, the wave becomes more localized around the point where $S$ cancels (black dashed line). The horizontal displacement is also larger, consistent with the analysis of \cite{leclerc2022}.}
    \label{fig:boosting}
\end{figure}

 We compute the linear normal modes of the solar interior with $S = S_{\odot}(r)$, $S = 10 S_{\odot}(r)$, and $S = 50 S_{\odot}(r)$, where $S_{\odot}(r)$ is derived from the calibrated 1D solar model used in the simulation. We compare the eigenfunctions of the Lamb-like wave at $\ell = 6.754$ in these three spectra. Figure~\ref{fig:boosting} shows the results, the blue curves being the solar model simulated in MUSIC. 

 As the values of $S$ increase, the Lamb-like waves become more confined around their cancellation point, and the horizontal displacement $\xi_\mathrm{h}$ dominates the vertical displacement $\xi_r$, as calculated in \cite{leclerc2022}. Although properties of $f$-modes at low $l$ are inherited from topology, the procedure above explains why these modes are additionally delocalized in the Sun. We expect topological modes in stars with steeper profiles of buoyant-acoustic frequencies to be closer to the academic situation studied in \cite{leclerc2022}.

\section{Amplitude of the modes at the surface and observability} 
\label{apdx:surf_amplitude}
\begin{figure}
    \centering
    \plotone{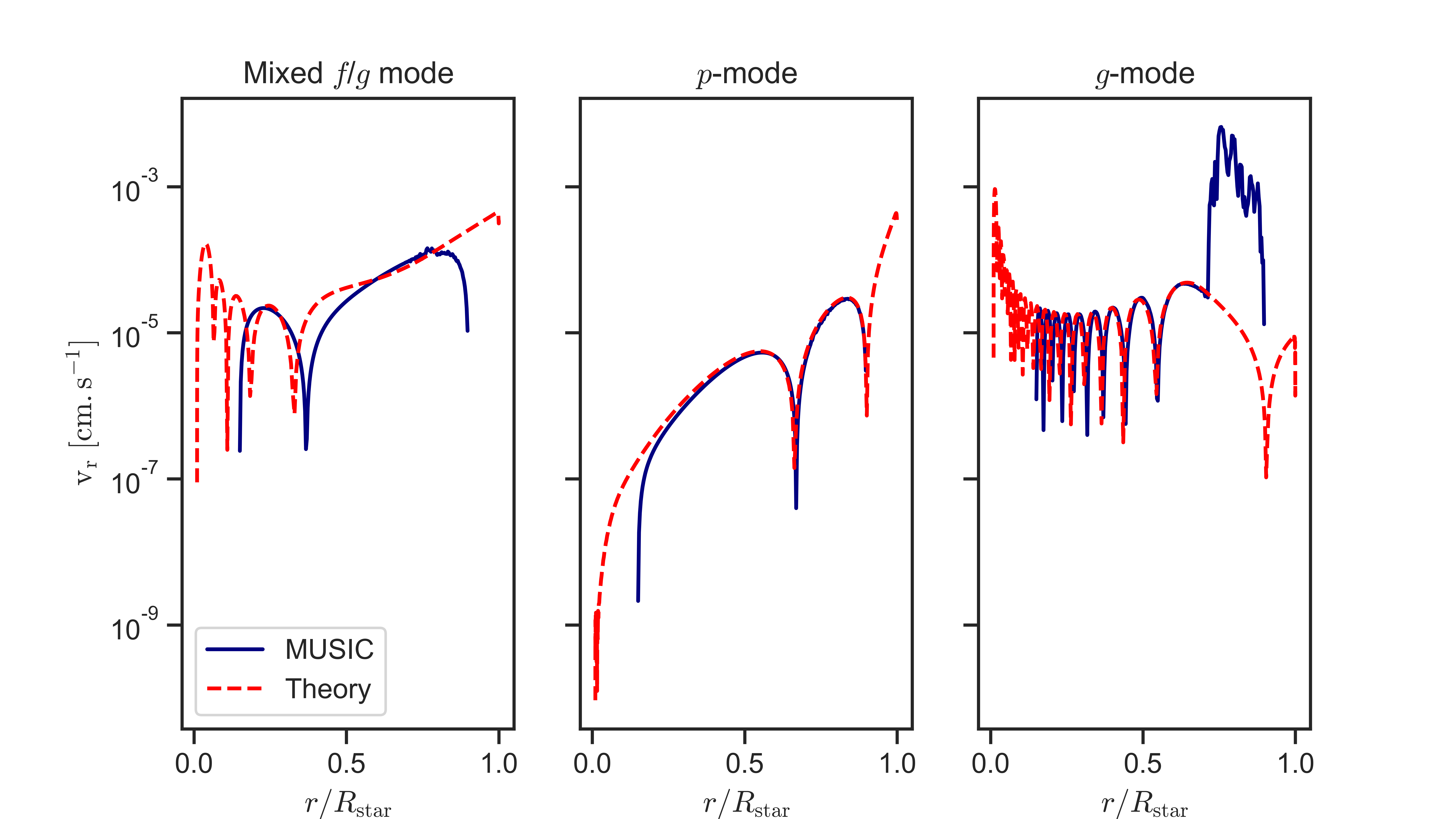}
    \caption{\textbf{Amplitude of standing modes} To estimate the surface amplitude of a mode, we compute its linear eigenfunction (dashed red lines) and fix its amplitude at $r = 0.6 \Rstar$ with the one measured in the MUSIC simulations (blue lines). We can then extrapolate the amplitude of the mode at $r = \Rstar$.}
    \label{fig:surf_amplitude}
\end{figure}

Due to numerical constraints, the radial domain of the simulations is restricted to $r = 0.15 \Rstar$ and $r = 0.9 \Rstar$.  Starting from data on this restricted domain, we identify normal modes excited by convective motion, and estimate their amplitude at the surface of the Sun. 
We begin by measuring the amplitude of the normal mode in the simulation at a selected radius within the radiative zone (outside the convective envelope and penetration zone to avoid measuring velocities from convective motions). This is done using the radial velocity extracted from the power spectrum. We choose $r= 0.6 \Rstar$, and verify that this choice does not affect the subsequent results. Next, we calculate the amplitudes of the linear modes for a non-truncated star to determine the ratio of the mode amplitude at  $r= 0.6 \Rstar$ to that at the surface. This correction factor is then applied to estimate the mode amplitude at the surface of the Sun. Due to the difference in cavity size between the simulation and the complete 1D model, the eigenfunctions and eigenfrequencies of the normal modes are not exactly identical. In particular, the mixed $f$/$g$-mode in the simulation is at 344 $\mu$Hz, the hybridization occurring with the $g_1$ mode, whereas in the extended 1D model and thus in the Sun, its frequency is 265 $\mu$Hz, the hybridization being with the $g_5$ mode. The left panel of Figure \ref{fig:surf_amplitude} shows the mixed modes for the simulation (blue line) and for the extended linear model (red dashed line).

Estimates for the surface amplitude of different modes are presented in the column $\velsurf$ of Table.~\ref{tab:modes_ampl}. We obtain an estimate of $2.1 \times 10^{-4} {\rm cm ~ s^{-1}}$ for the surface amplitude of the mixed $f$/$g$ mode, similar as $p$-modes in the same frequency range. However, in this frequency range, $g$-modes have significantly lower surface amplitudes. The only $g$-modes with similar amplitude as the mixed $f$/$g$ mode are at much lower frequencies, typically around $\sim$ 50 or 100 $\mu$Hz. 
Because stellar hydrodynamical simulations are far from stellar interior regimes, we find it more pertinent to compare the relative amplitudes between different modes rather than using the absolute amplitude of a given mode in $\rm cm ~ s^{-1}$.
This frequency dependence of the highest amplitude mode plays a key role in observational detection. In helioseismology, surface convection noise, which is the main obstacle to detecting normal modes, intensifies with decreasing frequencies \citep{Pincon2021}. Thus, despite having similar surface amplitudes, the mixed $f$/$g5$ mode presents a significantly higher likelihood of detection compared to $g$-modes.

A limitation of our study is that the convection zone is truncated at $r = 0.9 \Rstar$. However, it is well known that the outer layers of the convection zone play a major role in the excitation of the modes, at least for the high frequency ones. To test this, we run an additional simulation of the same solar model that extends from $r = 0.3 \Rstar$ to $r = 0.98 \Rstar$. We do not include  more convective layers for numerical reasons. The extension of the numerical domain to the photosphere ( $r = \Rstar$ ) is an open challenge for stellar hydrodynamical simulations given the sharp decrease of the pressure scale height with increasing radius. Because the size of the cavity is different, modes frequencies are shifted compared to our main simulations. We now identify the mixed $f$/$g$ mode at 252.8 $\mu$Hz.
The estimated surface amplitude using this simulation, $\velsurfext$, are presented in the last column of Table \ref{tab:modes_ampl}. The amplitudes of all modes increase significantly. This is expected as the inclusion of more external layers of the convective envelope drives convection stronger as was shown in \citet{Vlaykov2022}. Nevertheless, our conclusions regarding the relative amplitudes of the modes remain unchanged.

\begin{table}[t]
   \caption{\textbf{Surface amplitude $\velsurf$ of normal modes estimated using MUSIC simulation.} The angular degree and frequency of the normal modes correspond to those of the complete 1D model and are therefore consistent with the predictions for the Sun. See the text for details on the methodology used for this estimation. As there are no $g$-modes with a frequency of 230 $\mu$Hz in the simulation extended to $r=0.98 \Rstar$, we take instead the highest frequency $g$-mode for $\ell$ = 4, which is 177 $\mu$Hz.
   }
   \label{tab:modes_ampl}
   \centering
\begin{tabular}{c c c c c} 
     \hline \hline
      Mode  &  $\ell$ & $\omega / 2\pi$ ($\mu$Hz) & $\velsurf$ (cm.s$^{-1}$) & $\velsurfext$ (cm.s$^{-1}$) \\
      \hline
      Mixed $f$/$g$ & $4$ & 265 & $2.1 \times 10^{-4}$ & $3.3 \times 10^{-2}$\\
      $p$ & $4$ & 610 & $2.6 \times 10^{-4}$ & $4.3 \times 10^{-2}$\\
      $p$ & $2$ & 543 & $1.1 \times 10^{-4}$ & $2.4 \times 10^{-2}$\\
      $g$ & $4$ & 230 & $3.7 \times 10^{-5}$ & -\\
      $g$ & $4$ & 177 & - & $3.3 \times 10^{-3}$\\
      $g$ & $4$ & 93 & $5.3 \times 10^{-6}$ & $9.6 \times 10^{-3}$ \\
      $g$ & $2$ & 100 & $2.1 \times 10^{-5}$ & $1.7 \times 10^{-2}$\\
      $g$ & $4$ & 51 & $1.1 \times 10^{-4}$ & $3.7 \times 10^{-3}$\\
      \hline
   \end{tabular}
\end{table} 

Based on these results, an alternative estimate of the $f/g$ mode amplitude at the surface can be obtained by using the observed amplitudes for $p$-modes, and assuming that the ratio of these amplitudes to that of the $f/g$ mode is accurately described by linear theory. This method was already used in \citet{belkacem2022}. \citet{Davies2014} use observations of low-order, low-degree $p$-modes to estimate their amplitude as a function of frequency (see their Fig. 4). Using their results and assuming that the amplitude-frequency relationship remains valid at low frequencies, we estimate the surface amplitude of a $p$-mode with frequency $\sim 265 \mu$Hz, and therefore of the mixed $f$/$g$ mode, to be approximately 0.02 ${\rm cm ~ s^{-1}}$, which is similar to amplitude estimated using our simulation extended to $r = 0.98 \Rstar$. To date, the most accurate data available comes from the GOLF mission. With a time series spanning over 22 years, the detection threshold is approximately 0.3 ${\rm cm ~ s^{-1}}$ around 265 $\mu$Hz \citep{belkacem2022}.


\bibliography{scibib}{}
\bibliographystyle{aasjournal}



\end{document}